# On the hyperbolic Pennes equation description
# of the heat pulse-human   cornea interaction

J Marciak-Kozlowska [1], M.Kozlowski [2,*]

[1] Institute of Electron Technology, Warsaw, Poland
[2] Physics Department, Warsaw University, Warsaw, Poland

* Corresponding author,e-mail; miroslawkozlowski@aster.pl

Abstract

In this paper laser heating of human *cornea* is investigated. The new heat transport equation in *cornea* –hyperbolic Pennes equation is formulated and solved. It is shown that for ultra-short laser pulses the thermal energy propagates with finite velocity whereas for parabolic Pennes equation velocity is infinite. Moreover the thermal energy is concentrated in thermal wave front The model calculation are performed for *cornea* width l=950 $\mu$m, relaxation time $\tau = 20$ s and laser pulse duration $\Delta$ t=1 and 5 s

.

1. Introduction

The human cornea is a transparent, avascular, and highly specialized connective tissue that provides ~70 % of the total refraction in the optical system of the eye. Other essential properties of the cornea include a protection against noxious agents, biomechanical stability, and structural resiliency, as well as the ability to filter out damaging UV light thereby protecting both the crystalline lens and retina against injury. The human cornea (thickness ~530 µm) is a multi-layered tissue composed of five main layers: the epithelium (~50 µm), Bowman's layer (~10 µm), the stroma (~450 µm), Descemet's membrane (~5-15 µm), and the endothelium (~5 µm). In the healthy eye, these layers interact in a complex manner to strictly maintain the properties of the cornea. Increased biochemical knowledge of normal and diseased corneas are essential for the understanding of corneal homeostasis and pathophysiology.[1]

Basically, laser pulse can interact with biological tissue in five ways: (a) the electromechanical mode, (b) ablation, (c) photothermal (congulative and vaporizing) processes, (d) photochemical (photodynamics) reaction, (e) biostimulation and wound heating.

In this paper we will study in detail the photothermal processes. In the first approximation the cornea is the semiconductor and the electromagnetic field of the laser pulse generates the transport of charge, mass and heat in cornea. These transport phenomena in first approximation can be described as the diffusion processes

Knowledge on heat transfer in living tissues has been widely used in therapeutic applications. For further studying thermal behavior in biological bodies, many models describing bioheat transfer have been developed [1]. Due to simplicity and validity, the Pennes model is the most commonly used one among them. The applications of this relatively simple bioheat equation include simulations of hyperthermia and cryosurgery , thermal diagnostics .The Pennes bioheat equation describes the thermal behavior based on the classical Fourier's law. As is well known, Fourier's law depicts an infinitely fast propagation of thermal signal, obviously incompatible with physical reality.[2 Thus a modified flux model for the transfer processes with a finite speed wave is suggested [3] and solve the paradox occurred in the classical model. This thermal wave theory introduces a relaxation time that is required for a heat flux vector to respond to the thermal disturbance (that is, temperature gradient) as The relaxation time is approximated as $\tau = \alpha / V^2$. Here, $\alpha$ is the thermal diffusivity, and $V$ denotes the heat propagation velocity in the medium. In homogenous materials such as common metals, the relaxation time ranges from $10^{-8}$ to $10^{-14}$ s [3]. The heating processes are mostly much longer than this time scale. This is why the phenomenon of the heat wave is difficult to observe in homogenous substances. In reality, the living tissues are highly nonhomogenous, and accumulating enough energy to transfer to the nearest element would take time. The literatures [4,5] and many others reported the value of $\tau$ in biological bodies to be 20–30 s. Mitra et al. [4] found the relaxation time for processed meat is of the order of 15 s. Recently, Roetzel et al. [5] experimentally investigated the relaxation thermal behavior in nonhomogenous  materials

2.  Eye surgery with ultra-short laser pulses

 One of the earliest and most successful application of short laser pulses is laser eye surgery. To correct  near  or farsightandness in a person portion of the cornea are cut and reshaped   so that  the cornea will then correctly focus. In photorefractive kertectomy UV laser light is used to photoablate and reshape of  cornea but also ablates the protective endothelial layers on the corneal surface. In a modification of this procedure laser assisted in situ keratomileusis {LASIK}, the corneal tissue is first exposed by cutting a flap from the surface layer with a mechanical blade. Then UV light is used to photoablate the corneal tissue and the flap  is replaced. The preservation of the surface endothelial layer helps speed recovery from the surgery. While UV laser ablation via linear absorption was a great improvement other mechanical cutting of the corneal tissue , new technique using femtosecond laser ablation offer further improvements  in the surgical outcome.

Juhasz et al. [6] developed an optical technique to cut this flap by taking advantage of the nonlinearity of ablation by femtosecond laser. The laser based cutting of the flap  is far more precise and is  now  performed at many clinics and offers significant reduction in side effects [2]
 The penetration of laser thermal energy in *cornea* strongly depends on the underlaing mechanism. In this paper we describe new mode of cornea heat transport i.e. thermal waves

When exposed to attosecond laser pulses cornea behaves as the multi-layered structure. On the edges of the layers the thermal waves are reflected. Moreover the thermal energy is focused on the front of the thermal wave and can be localized. In the diffusion mode[3] the thermal energy is smoothly distributed over all heating region. The localization of the energy in front of the thermal wave enables the control of the heating processes

3   Mathematical formulation

High - energy beams like X - rays and lasers are being increasingly used in a variety of material processing, manufacturing and biomedical applications. Recently developed short pulses lasers, e.g. femtosecond and attosecond lasers have the additional ability to investigate the matter on the atomic level [3].

The non - zero value of the relaxation time changes the mathematical structure of the transport equation. The parabolic partial differential equations like: Fick law, Fourier law are not valid when $\tau \neq 0$. In that case, more general hyperbolic transport equation like non - Fourier, Heaviside equation are applied to the study of transport processes [3].

According to the constitutive relation in the non - Fourier processes, heat, mass or charge flux $q$ obeys the relation [3]:

$$\vec{q}_T(r, t + \tau_T) = -k\nabla T \tag{1}$$

for thermal processes characterized by temperature T;

$$\vec{q}_n(r, t + \tau_n) = -D\nabla n \tag{2}$$

for diffusion of species characterized by density $n$.

In formulae (1) and (2) $k$ and $D$ denotes the thermal conductivity and diffusion coefficient respectively. The $\tau_{n,T}$ denotes the relaxation time for density and heat diffusion respectively. The gradients $\nabla T$ and $\nabla n$ established in the material at the time $t$ result in a flux that occurred at later time $t + \tau$ due to the insufficient time of response.

The Taylor's series expansion when applied to the formulae (I.1) and (I.2) gives

$$\vec{q}_{n,T}(\vec{r}, t) + \tau_{n,T}\frac{\partial \vec{q}_{n,T}(\vec{r}, t)}{\partial t} + \frac{\partial^2 \vec{q}_{n,T}(\vec{r}, t)}{\partial t^2}\frac{\tau_{n,T}^2}{2} = -B_{n,T}\nabla\left(\begin{array}{c}n(\vec{r}, t)\\T(\vec{r}, t)\end{array}\right), \tag{3}$$

where $B_n = D$, $B_T = k$.

In the linearized theory of the transport phenomena the time lag is assumed to be small and the higher order terms in (I.3) are neglected. By retaining only the first – order term in $\tau$, formula (I.3) can be written as

$$\vec{q}_{n,T}(\vec{r},t) + \tau_{n,T}\frac{\partial \vec{q}_{n,T}(\vec{r},t)}{\partial t} = -B_{n,T}\nabla\begin{pmatrix} n(\vec{r},t) \\ T(\vec{r},t) \end{pmatrix}. \tag{4}$$

Combining equations (I.4) with conservations laws the hyperbolic transport equations for $n$, $T$ can be obtained [1, 2].

As a simple illustration of the hyperbolic transport processes let us consider the model independent hyperbolic equation, Heaviside equation*:

$$\varepsilon\frac{\partial^2 u}{\partial t^2} - c^2\frac{\partial^2 u}{\partial x^2} + \frac{\partial u}{\partial t} = 0, \tag{5}$$

where $\varepsilon > 0$; if we set $\varepsilon = 0$, (I.5) formally reduces to the parabolic equation. The hyperbolic equations of the type (I.5) are used in physics, biophysics and archaeology**.

We might expect that as $\varepsilon \to 0$ the causal fundamental solution of (.5) corresponding to a source point at $P_0(x_0, t_0)$ reduces to the causal fundamental solution of the parabolic equation. Indeed the causality condition for both problems requires that $u(x,t) = 0$ for $t < t_0$. Furthermore the causal fundamental solution of (I.5) vanishes outside the forward characteristic sector $|x - x_0| < \hat{c}(t - t_0)$, where $\hat{c} = \dfrac{c}{\sqrt{\varepsilon}}$ . As $\varepsilon \to 0$ the boundary $\sqrt{\varepsilon}|x - x_0| = c(t - t_0)$ of this sector tends to the line $t = t_0$ which is a characteristic line for the parabolic equation. In the limit therefore the fundamental solution of (I.5) is expected to be non - zero in the region $t > t_0$, a result appropriate for the parabolic equation.

Let

$$u(x,t) = \exp\left[-\frac{1}{2\varepsilon}(t - t_0)\right]\upsilon(x,t) \tag{6}$$

in (I.5) and we obtain

---

* O.Heaviside, Electrical Papers 1876. There have been several reprints of this book over years, for example Chelsea Publishing Company , New York 1970.

** M. A. Pelc, *Time delayed processes in physics, biophysics and archaeology,* arXiv:0706.0011.

$$\varepsilon \frac{\partial^2 \upsilon}{\partial t^2} - c^2 \frac{\partial^2 \upsilon}{\partial x^2} - \frac{1}{4\varepsilon} \upsilon = 0 . \tag{7}$$

The solution of (I.7) has the form

$$u(x,t) = \frac{1}{\sqrt{4c^2\varepsilon}} \exp\left[-\frac{(t-t_0)}{2\varepsilon}\right] I_0\left[\frac{1}{\sqrt{4c^2\varepsilon}} \sqrt{\frac{c^2}{\varepsilon}(t-t_0)^2 - (x-x_0)^2}\right]$$

$$\text{for } |x-x_0| < \frac{c}{\sqrt{\varepsilon}}(t-t_0) \tag{8}$$

$$u(x,t) = 0 \qquad\qquad \text{for } |x-x_0| > \frac{c}{\sqrt{\varepsilon}}(t-t_0)$$

As $z \to \infty$, the modified Bessel function $I_0(z)$ has the asymptotic behavior

$$I_0(z) \approx \frac{1}{\sqrt{2\pi\, z}} e^z, \qquad z \to \infty . \tag{9}$$

Also for $t > t_0$ and $0 < \varepsilon << 1$ we have

$$\sqrt{\frac{c^2}{\varepsilon}(t-t_0)^2 - (x-x_0)^2} = \frac{c}{\sqrt{\varepsilon}}(t-t_0)\left[1 - \frac{\varepsilon}{2c^2}\frac{(x-x_0)^2}{(t-t_0)} + ...\right]. \tag{10}$$

Thus

$$\left(4c^2\varepsilon\right)^{\frac{1}{2}} \sqrt{\frac{c^2}{\varepsilon}(t-t_0)^2 - (x-x_0)^2} = \frac{t-t_0}{2\varepsilon} - \frac{(x-x_0)^2}{4c^2(t-t_0)} + ... \tag{11}$$

As $\varepsilon \to 0$ in (I.8) the argument of $I_0$ tends to infinity. Using (I.9), (I.10) and (I.11) we obtain in the limit as $\varepsilon \to 0$

$$u(x,t) = \frac{1}{\sqrt{4\pi c^2(t-t_0)}} \exp\left[-\frac{(x-x_0)^2}{4c^2(t-t_0)}\right] \qquad \text{for } t > t_0$$

$$u(x,t) = 0 \qquad\qquad\qquad \text{for } t < t_0 \tag{12}$$

which is identical to the fundamental solution of the parabolic equation.

We conclude that for $\varepsilon \neq 0$ the Heaviside equation (I.5) is the generalized diffusion equation. The parameter $\varepsilon$ has the dimension of time and is recognized as the relaxation time for the processes described by the Heaviside equation. Considering that in all realistic transport phenomena $\varepsilon \neq 0$ ***, the Heaviside

---

*** M. Kozlowski, J. Marciak - Kozlowska, *From femto- to attoscience and beyond,* Nova science Publishers, 2009,USA

equation and more general Proca, Klein – Gordon equation are the master equation for those processes.

## 3. Hyperbolic versus parabolic

In the description of the evolution of any physical system, it is mandatory to evaluate, as accurately as possible, the order of magnitude of different characteristic time scales, since their relationship with the time scale of observation (the time during which we assume our description of the system to be valid) will determine, along with the relevant equations, the evolution pattern. Take a forced damped harmonic oscillator and consider its motion on a time scale much larger than both the damping time and the period of the forced oscillation. Then, what one observes is just a harmonic motion. Had we observed the system on a time scale of the order of (or smaller) than the damping time, the transient regime would have become apparent. This is rather general and of a very relevant interest when dealing with dissipative systems. It is our purpose here, by means of examples and arguments related to a wide class of phenomena, to emphasize the convenience of resorting to hyperbolic theories when dissipative processes, either outside the steady-state regime or when the observation time is of the order or shorter than some characteristic time of the system, are under consideration. Furthermore, as it will be mentioned below, transient phenomena may affect the way in which the system leaves the equilibrium, thereby affecting the future of the system even for time scales much larger than the relaxation time.

Parabolic theories of dissipative phenomena have long and a venerable history and proved very useful especially in the steady-state regime [1.1]. They exhibit however some undesirable features, such as acausality (see e.g., [1.2, 1.3]), that prompted the formulation of hyperbolic theories of dissipation to get rid of them. This was achieved at the price of extending the set of field variables by including the dissipative fluxes (heat current, non-equilibrium stresses and so on) at the same footing as the classical ones (energy densities, equilibrium pressures, etc), thereby giving rise to a set of more physically satisfactory (as they much better conform with experiments) but involved theories from the mathematical point of view.

These theories have the additional advantage of being backed by statistical fluctuation theory, kinetic theory of gases (Grad's 13-moment approximation), information theory and correlated random walks (at least in the version of Jou *et al.)* [1.3].

A key quantity in these theories is the relaxation time $\tau$ of the corresponding dissipative process. This positive-definite quantity has a distinct physical meaning, namely the time taken by the system to return spontaneously to the steady state (whether of thermodynamic equilibrium or not) after it has been suddenly removed from it. It is, however, connected to the mean collision time $t_c$ of the particles responsible for the dissipative process It is therefore appropriate to interpret the relaxation time as the time taken by the corresponding dissipative flow to relax to its steady value. Thus, it is well known that the classical Fourier law for heat current, [1.3]

$$\vec{q} = -\kappa \, \nabla T \tag{13}$$

with $\kappa$ the heat conductivity of the fluid, leads to a parabolic equation for temperature (diffusion equation)

$$\frac{\partial T}{\partial t} = \chi \, \nabla^2 T, \qquad \chi = \frac{\kappa}{\rho \, c_V} = \frac{\kappa}{C_V}; \qquad C_V = \rho \, c_V. \tag{14}$$

(where $\chi$, $\rho$ and $c_V$ are diffusivity, density and specific heat at constant volume, respectively), which does not forecast propagation of perturbations along characteristic causal light-cones, that is to say, perturbations propagate with infinite speed. This non-causal behavior is easily visualized by taking a look at the thermal conduction in an infinite one dimensional medium. Assuming that the temperature of the line is zero for $t < 0$, and putting a heat source at $x = x_0$ when $t = 0$, the temperature profile for $t > 0$ is given by

$$T \sim \frac{1}{\sqrt{t}} \exp\left[ -\frac{(x - x_0)^2}{t} \right] \tag{15}$$

implying that for $t = 0 \Rightarrow T = \delta(x - x_0)$, and for $t = t_1 > 0 \Rightarrow T \neq 0$. In other words, the presence of a heat source at $x_0$ is instantaneously felt by all observers on the line, no matter how far away from $x_0$ they happen to be. The origin of this behavior can be traced to the parabolic character of Fourier's law, which implies that the heat flow starts (vanishes) simultaneously with the appearance (disappearance) of a temperature gradient. Although $\tau$ is very small for phonon-electron, and phonon-phonon interaction at room temperature ($10^{-11}$ and $10^{-13}$ seconds, respectively), neglecting it is the source of difficulties, and in some cases a bad approximation as for example in superfluid Helium, and degenerate stars where thermal conduction is dominated by electrons -see [1.3], for further examples.

In order to overcome this problem Cattaneo and (independently) Vernotte by using the relaxation time approximation to Boltzmann equation for a simple gas derived a generalization of Fourier's law, namely [1.3]

$$\tau \frac{\partial \vec{q}}{\partial t} + \vec{q} = -\kappa \nabla T \ . \tag{16}$$

This expression (known as Cattaneo-Vernotte's equation) leads to a hyperbolic equation for the temperature (Heaviside equation) which describes the propagation of thermal signals with a finite speed

$$\upsilon = \sqrt{\frac{\chi}{\tau}} \ . \tag{17}$$

This diverges only if the unphysical assumption of setting $\tau$ to zero is made.

It is worth mentioning that a simple random walk analysis of transport processes naturally leads to Heaviside equation, not to the diffusion equation -see e.g. [1.4]. Again, the latter is obtained only if one neglects the second derivative term.

It is instructive to write (1.4) in the equivalent integral form

$$\vec{q} = -\frac{\chi}{\tau} \int_{-\infty}^{t} \exp\left[-\frac{(t-t')}{\tau}\right] \cdot \nabla T(x',t') dt' \ , \tag{18}$$

which is a particular case of the more general expression

$$\vec{q} = -\int_{-\infty}^{t} K(t-t') \nabla T(x',t') dt' \ . \tag{19}$$

The physical meaning of the kernel $K(t-t')$ becomes obvious by observing that

for $\quad K = \kappa\delta(t-t') \rightarrow \vec{q} = -\kappa\nabla T \qquad$ (Fourier) $\qquad\qquad$ (20)

for $\quad K = \text{constant} \rightarrow \dfrac{\partial^2 T}{\partial t^2} = \chi\nabla^2 T \qquad$ (wave motion) $\qquad\qquad$ (21)

i.e., $K$ describes the thermal memory of the material by assigning different weights to temperature gradients at different moments in the past. The Fourier law corresponds to a zero-memory material (the only relevant temperature gradient is the "last" one, i.e., the one simultaneous with the appearance of $q$). By contrast the infinite memory case (with $K = \text{constant}$) leads to an undamped wave. Somewhere in the middle is the Cattaneo-Vernotte equation, for which all temperature gradients contribute to $q$, but their relevance goes down as we move to the past.

As the third case, "intermediate memory" will be considered:

$$K(t-t') = \frac{K_3}{\tau}\exp\left[-\frac{(t-t')}{\tau}\right],\qquad\qquad (22)$$

where $\tau$ is the relaxation time of thermal processes. Combining Eqs. (2.13) and (2.2) we obtain

$$c_V\frac{\partial^2 T}{\partial t^2} + \frac{c_V}{\tau}\frac{\partial T}{\partial t} = \frac{K_3}{\rho\tau}\nabla^2 T \qquad\qquad (23)$$

and

$$K_3 = D_T c_V \rho. \qquad\qquad (24)$$

Thus, finally,

$$\frac{\partial^2 T}{\partial t^2} + \frac{1}{\tau}\frac{\partial T}{\partial t} = \frac{D_T}{\tau}\nabla^2 T. \qquad\qquad (25)$$

$$D_T = v^2\tau$$

In formula (2.16) v is the velocity of heat propagation and τ is the thermal relaxation time  From these comments it should be clear that different classes of dissipative systems may be described by different kernels. Obviously, when studying transient regimes, i.e., the evolution from a steady-state situation to a new one, $\tau$ cannot be neglected. In fact, leaving aside that parabolic theories are necessarily non-causal, it is obvious that whenever the time scale of the problem under consideration becomes of the order of (or smaller) than the relaxation time, the latter cannot be ignored. It is common sense what is at stake here: neglecting the relaxation time amounts – in this situation – to disregarding the whole problem under consideration. According to a basic assumption underlying the disposal of hyperbolic dissipative theories, dissipative processes with relaxation times comparable to the characteristic time of the system are out of the hydrodynamic regime.

### 3. Heat transport in human cornea

For medical and biological sciences the parabolic heat diffusion equation was developed and solved by H H Pennes [ 5 ]. In this paper we showed the general method for introducing the generalised hyperbolic Pennes equation (2.16) .In sequel part of the paper we present the solution of the hyperbolic Pennes equation for the : "step" boundary condition[3]:

$T(0,t)=T_0$             for $\Delta t = 1, 5$ s

$T(0,t) =0$             for $t > \Delta t$           ( 26 )

Considering the results of the paper [6] we use for the relaxation time in human cornea, $\tau = 20$s . and cornea thickness l= 950 $\mu$ m In Figs 1-8 the solution of the Eq.( 2.16) for different values of v, and $d$ were presented In Figs.1-4 $a$ the solutions of the hyperbolic Pennes equation (2.16 ) , for $\Delta t =1$s and v= 30 , 100, 300, ,1300 $\mu$ m/s respetively. In Figs 1-4 b the solution of the parabolic Pennes equation\ )were presented. In Figs 5-8 a,b the solutions of the hyperbolic and parabolic Pennes equations were presented for $\Delta t=$ 5s.

From the structure of the hyperbolic Pennes equation we conclude that the wave mode is predominant for low values of heat velocity propagation. In Figs 1-8

a,b the wave mode can be observed clearly for v= 30 $\mu$ m/s. The diffusion mode is dominant for v= 1300 $\mu$ m/s and in that case the solutions of the hyperbolic and parabolic Pennes equation are the same.

For the future eyes surgery it is important to note that for thermal pulses with $\Delta$t smaller than $\tau$ the new mode -thermal waves can be generated. Among the many surgical procedures in which laser heating is employed the cornea laser heating , laser thermo-keratoplasty ( LTK) is very important[8]. In paper [2] the Fourier heat transport equation-Pennes parabolic equation is applied to the study of the thermal processes in LTK. As was shown above the application of the hyperbolic Pennes equation offersthe new possibility in comparison to Fourier based models: (i) the velocity of heat propagation is finite, and (ii) heat energy is concentrated in the front of the thermal wave

References

1.  Peng Q. et al., *Rep. Prog. Phys.* 71 (2008) 056701.

2.  Kozlowski M., Marciak – Kozlowska J., *Thermal Processes Using Attosecond Laser Pulses*, Springer, USA 2006.

3.  Polyanin A. D., *Handbook of linear partial differential equations for engineers and scientists*, Chapman and Hall/CRC, London, 2005.

4.  Franco S., Almeida J. B., Parafita M., Corneal thickness and elevation maps computed from optical rotary scans. *Journal of Refractive Surgery* Vol. 20 *No. 5* September/October 2004.

5   H.H. Pennes. Analysis of tissue and arterial blood temperatures in the resting human forearm. *Journal of Applied Physiology*, 1(2):93–122, 1948.

Figure captions

Figs.1-4 *a* The solutions of the hyperbolic Pennes equation (2.16 ), for $\Delta t = 1s$ and v= 30 , 100, 300, ,1300 $\mu$ m. Figs 1-4 b the solution of the parabolic Pennes equation

Figs 5-8 a,b the solutions of the hyperbolic and parabolic Pennes equations for $\Delta t$= 5s and v= 30 , 100, 300, ,1300 $\mu$ m.

Out[87]=

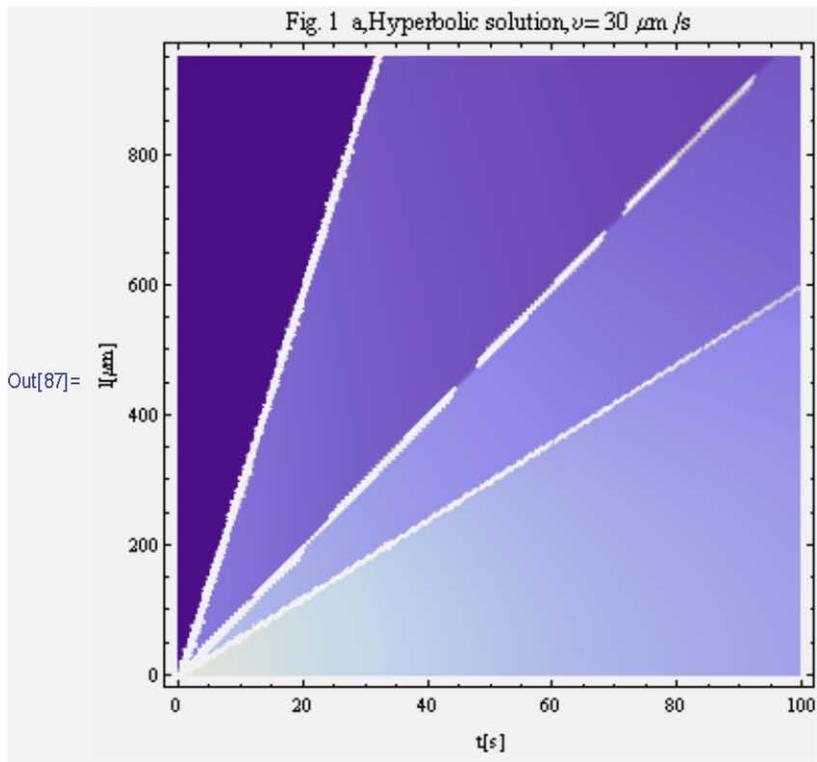

Fig. 1  a,Hyperbolic solution,$\upsilon = 30\ \mu m\ /s$

Out[88]=

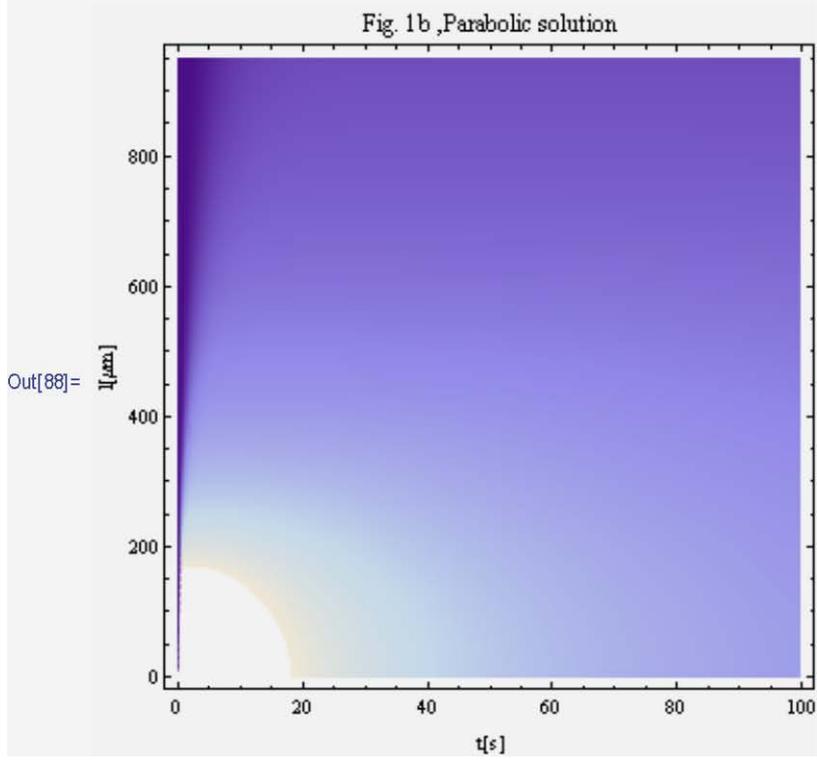

Fig. 1b ,Parabolic solution

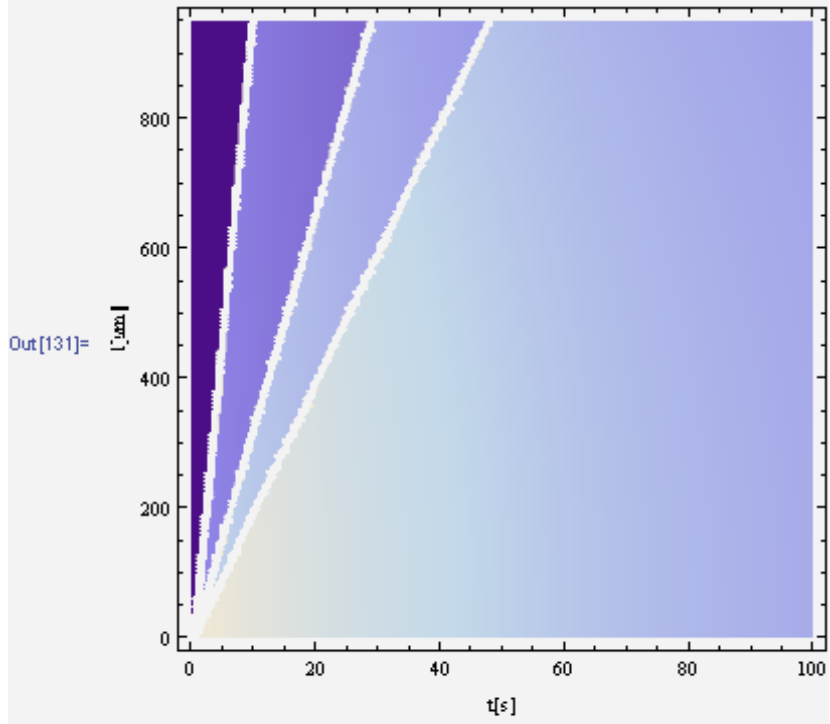

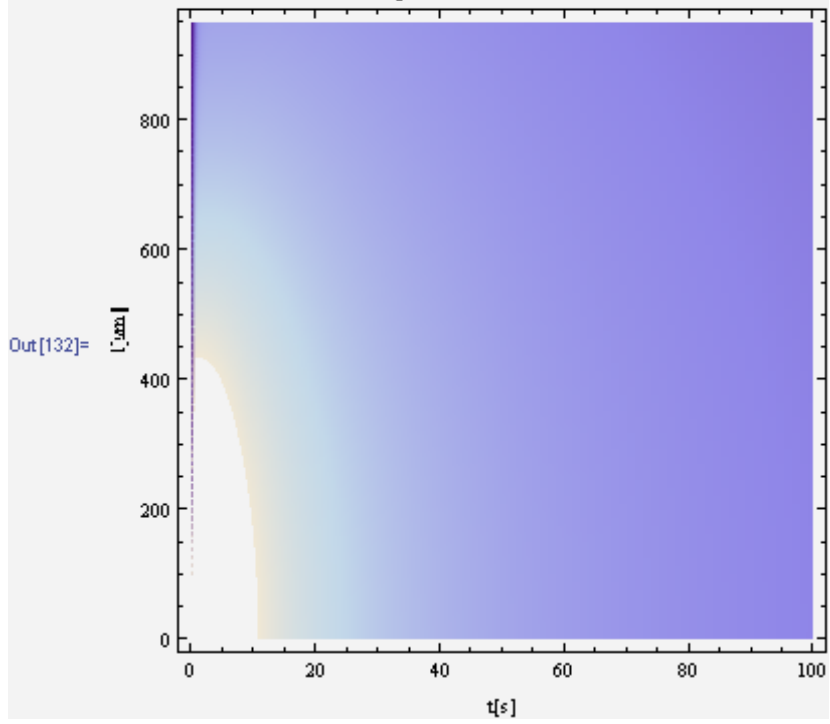

Out[131]=

Out[132]=

Out[153]=

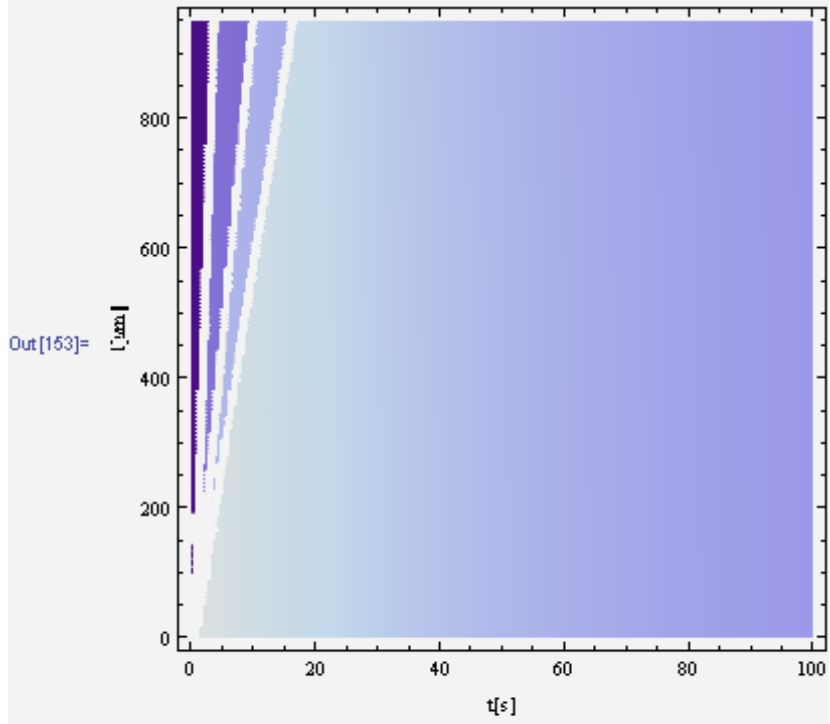

Out[154]=

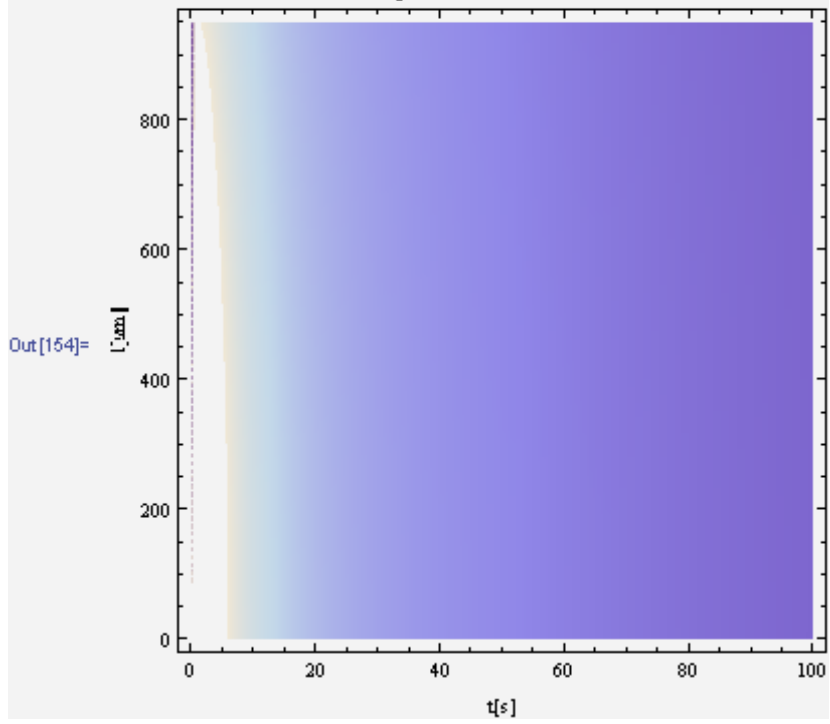

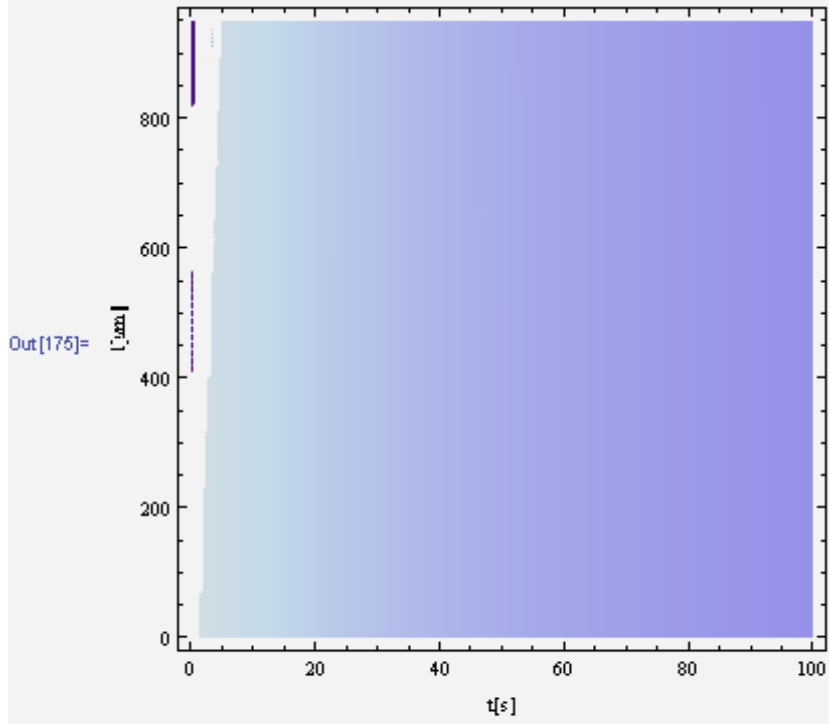

Out[175]=

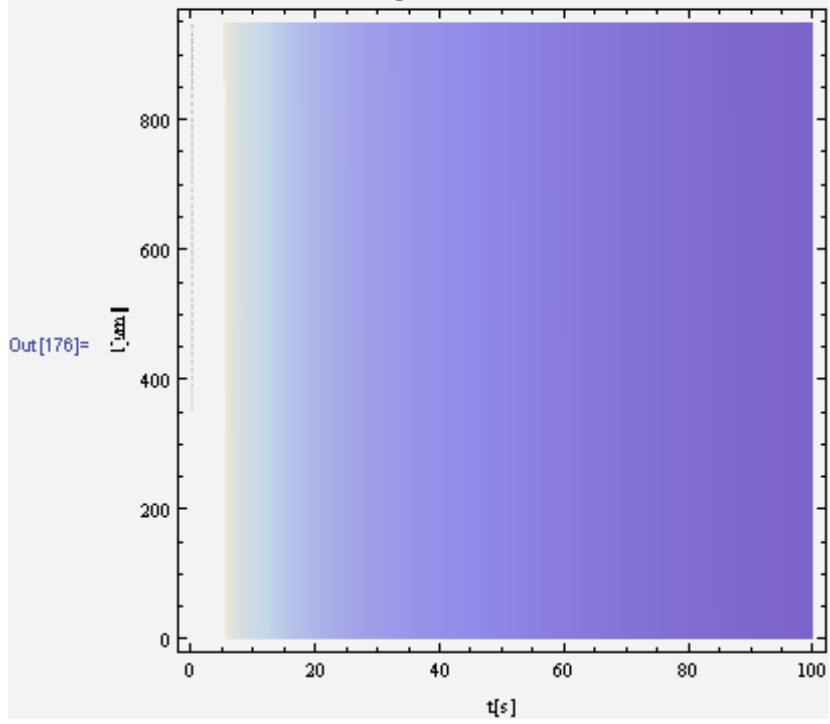

Out[176]=

Out[198]=
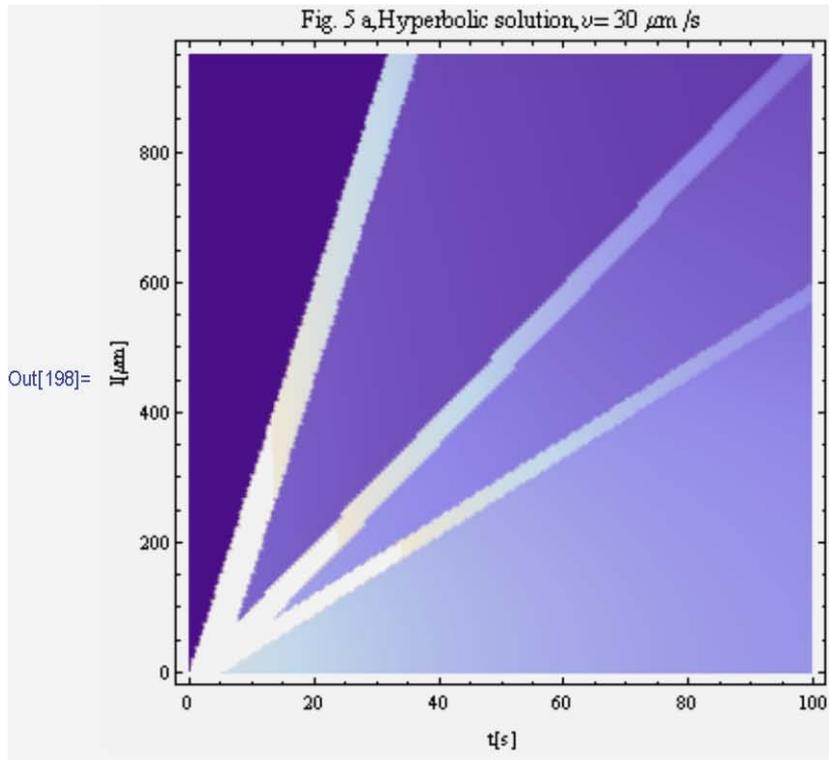
Fig. 5 a,Hyperbolic solution,$\upsilon = 30 \ \mu m \ /s$

Out[199]=
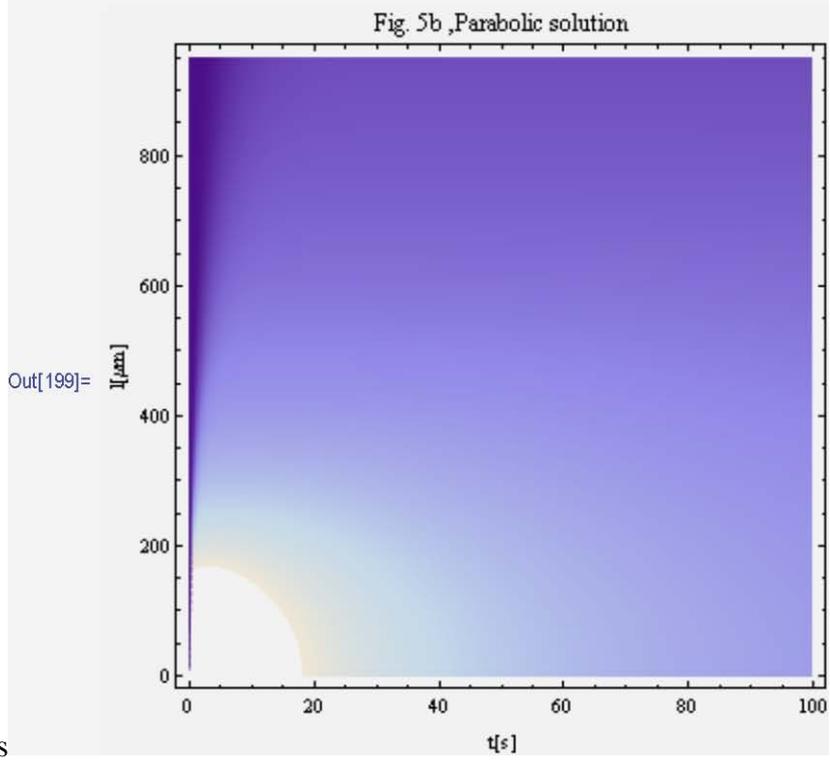
Fig. 5b ,Parabolic solution

s

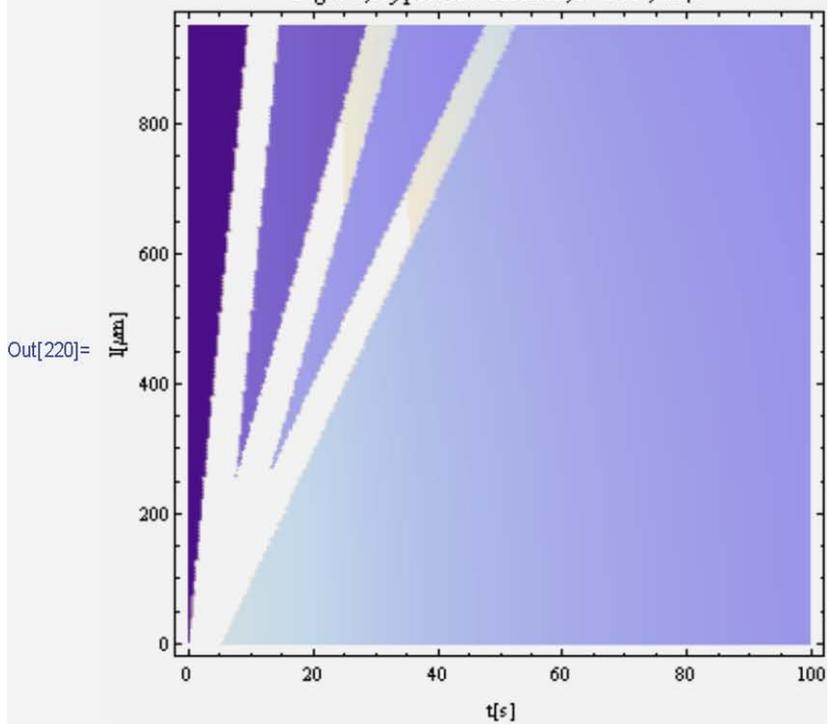

Out[220]=

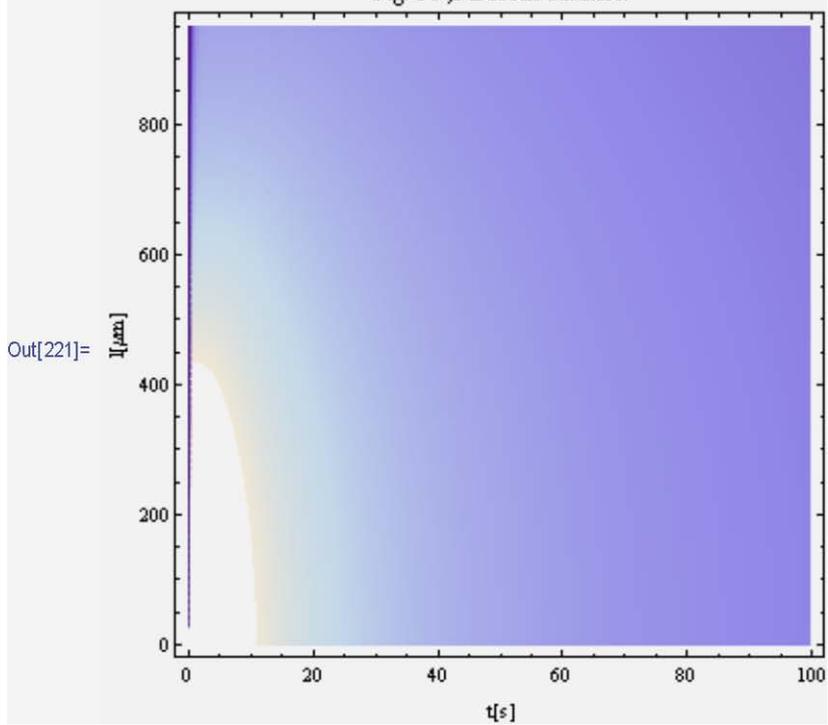

Out[221]=

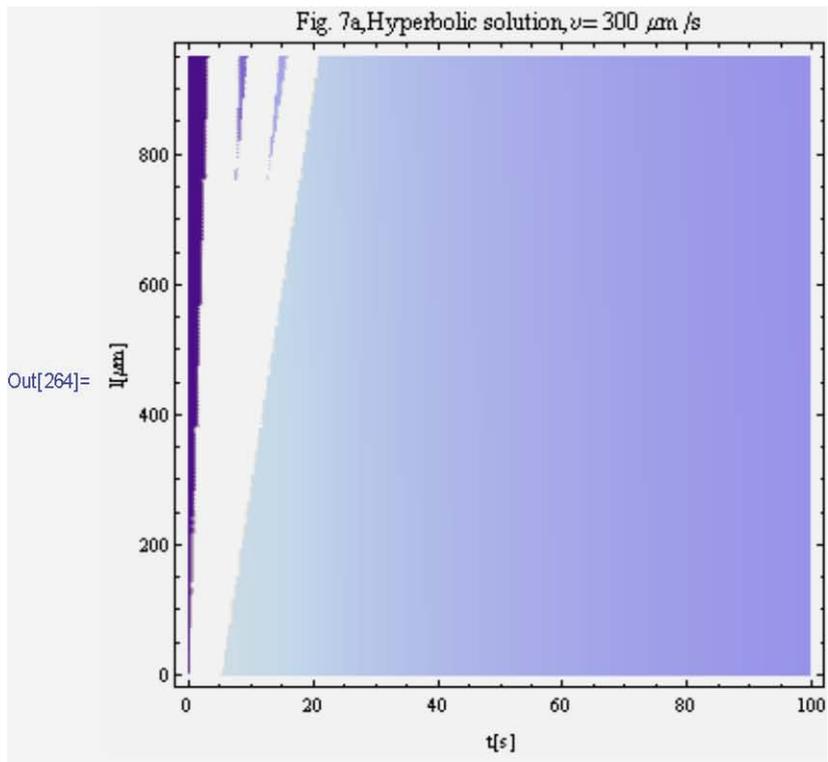

Out[264]=

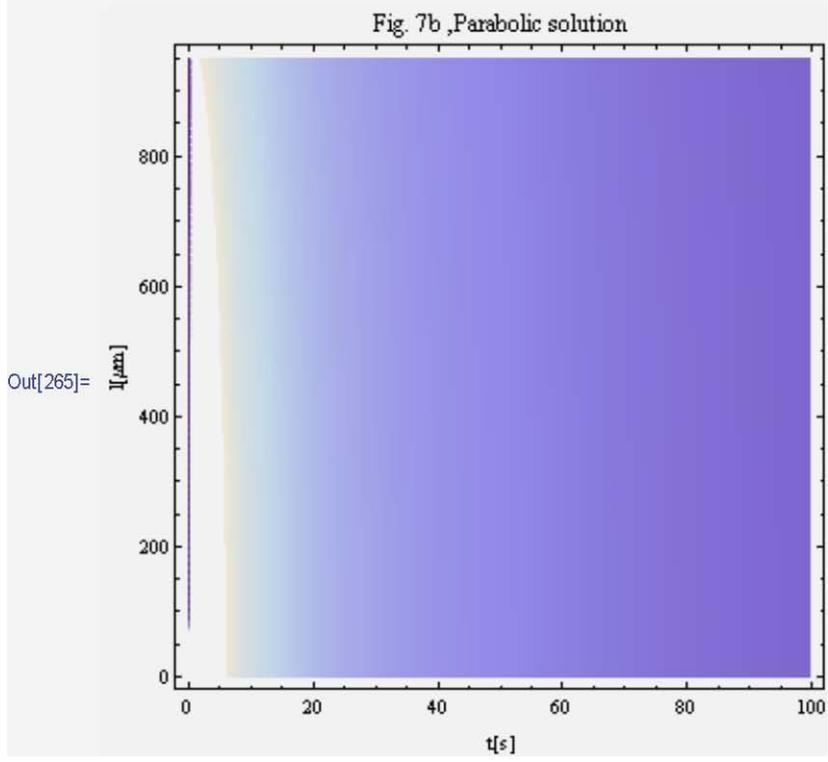

Out[265]=

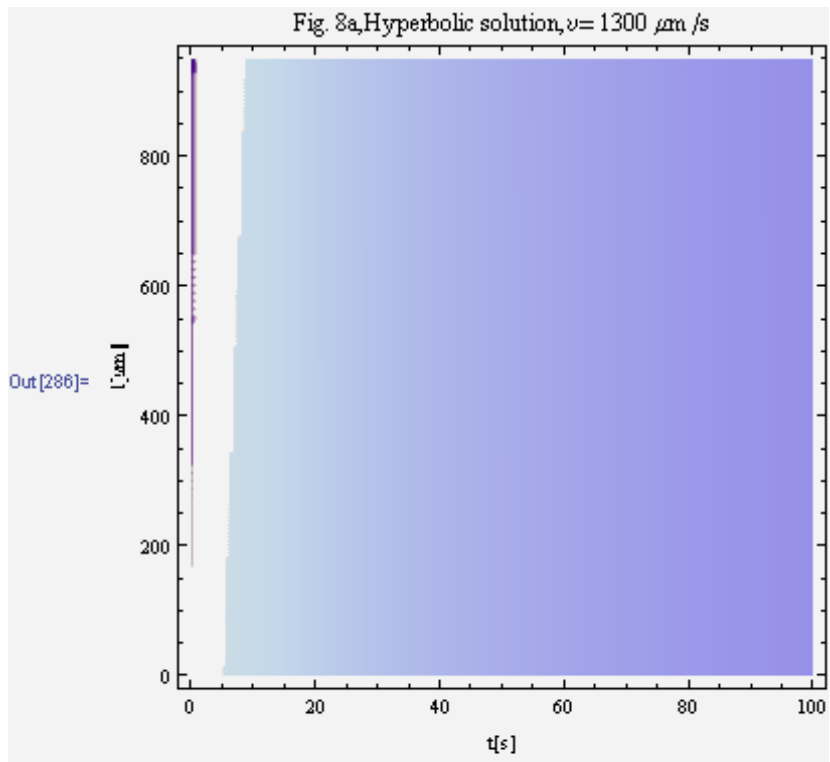

Out[286]=

Fig. 8a,Hyperbolic solution,$v = 1300\ \mu m\ /s$

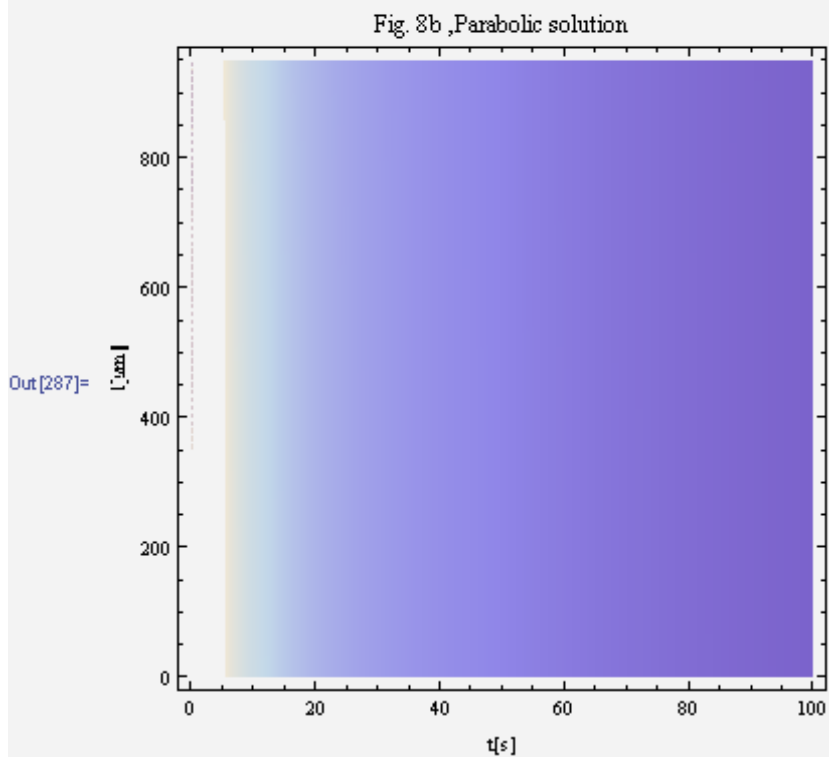

Out[287]=

Fig. 8b ,Parabolic solution

References

1.    Jou  D.,  Casas  –  Vázquez  J.,  Lebon  G.,  *Extended  Irreversible Thermodynamics*, Springer 2001.

2.    Tzou D. X., *Macro- To Micro-Scale Heat Transfer: Past, Present and Future,* Taylor and Francis, USA 1996.

3.    Kozlowski  M.,  Marciak  –  Kozlowska  J.,  *Thermal  Processes  Using Attosecond Laser Pulses*, Springer 2006.